\documentclass[a4paper,10pt]{article}
\usepackage{graphicx}
\usepackage{comment}
\usepackage{amssymb}
\usepackage{amsmath}
\usepackage{nicefrac}
\usepackage{graphicx}
\usepackage{dcolumn}
\usepackage{bm}
\usepackage{comment}

\begin{document}

\huge

\begin{center}
Opacity calculations for stellar astrophysics
\end{center}

\vspace{0.5cm}

\large

\begin{center}
Jean-Christophe Pain\footnote{jean-christophe.pain@cea.fr} and Franck Gilleron
\end{center}

\normalsize

\begin{center}
\it CEA, DAM, DIF, F-91297 Arpajon, France
\end{center}

\vspace{0.5cm}

\begin{abstract}
Opacity is a key ingredient of stellar structure and evolution. In the present work, we recall the role of opacity in asteroseismology, focusing mainly on two kinds of astrophysical objects: $\beta$ Cephei-type stars, and the Sun. The detailed opacity code SCO-RCG for local-thermodynamic-equilibrium plasmas is described and interpretation of laser and Z-pinch experiments are presented and discussed. The possible role of multi-photon processes on radiative accelerations is outlined, and the main aspects of opacity modeling which should be improved are mentioned.
\end{abstract}

\section{Introduction}\label{sec1}

The knowledge of opacity (frequency-dependent photo-absorption cross-section per mass unit) is crucial for the understanding and modeling of radiation transport. The applications encompass inertial confinement fusion, on the National Ignition Facility (NIF) or Laser M\'egajoule (LMJ), magnetic confinement fusion on International Thermonuclear Experimental Reactor (ITER) especially for the radiative losses at tungsten-coated tiles of divertor, and astrophysics. Models of stellar structure and evolution are very sensitive to radiative transfer and opacity. In section \ref{sec2}, we recall some basic concepts of opacity and asteroseismology and present the main sources of uncertainty in the modeling of main-sequence pulsators as well as the role of opacity in the ``kappa'' mechanism taking place in $\beta$ Cephei-type stars due to the opacity bump of the iron-group elements. In section \ref{sec3}, our opacity code SCO-RCG is briefly presented, and values of Planck and Rosseland mean opacities are discussed and compared to other models. In section \ref{sec4}, interpretations of laser and Z-pinch experiments are presented, and in section \ref{sec5}, the issue of the opacity in the interior of the Sun is raised together with the enigmatic iron transmission spectrum measured on the Sandia Z machine. The role of multi-photon processes on radiative accelerations is briefly discussed in section \ref{sec6}.

\section{Opacities and asteroseismology}\label{sec2}

\subsection{Historical considerations about radiative opacity}\label{sec2a}

In 1924, Payne-Gaposchkin demonstrates the great preponderance of hydrogen and helium in stars and in 1929, Russell publishes the first quantitative analysis of the chemical composition of the solar atmosphere \cite{Payne25,Russell29}. Opacity is already known as a key parameter of stellar models since 1926 \cite{Eddington26}. Although in such conditions, atoms are not fully ionized, photo-excitation and photo-ionization are neglected at that time. In the early sixties, Cox and Huebner introduce the two latter processes in the calculation of stellar opacities. Simon raises in 1982 the problem of the pulsation of Cepheids (named after $\delta$ Cephei) and its connexion to opacity of elements with atomic number $Z\geq$ 6. In the 1990s, two important projects come up: OPAL (LLNL) \cite{Iglesias96} and OP (Opacity Project, an international academic collaboration) \cite{OpacityProject,Seaton87} which provide the first generation of stellar opacity tables.

\subsection{Sources of uncertainty in the modeling of main-sequence pulsators}\label{sec2b}

Interior structure and temporal evolution of main-sequence pulsators are not satisfactorily understood. For massive stars ($M\ge M_{\odot}$), there are different sources of uncertainty. Among them, some important ones can be identified from the following considerations:

\vspace{5mm}

(i) Rotation reduces luminosity and internal temperature, increases density, and induces internal mixing yielding surface abundance changes during evolution.

\vspace{5mm}

(ii) Angular-momentum transport from the core to the envelope is needed to avoid that iron cores reach critical rotation in later evolutionary stages.

\vspace{5mm}

(iii) Mixing of material into the hydrogen-burning core (convection overshooting) affects the main-sequence lifetimes.

\vspace{5mm}

(iv) With the new solar abundances, an increase in the heavy-element opacities in the Sun is required to match helioseismic data. The same conclusion holds for pulsational mode excitation in massive stars. In sections \ref{sec2c} and \ref{sec2d}, we recall some basics of asteroseismology and its connection to opacity, through the so-called $\kappa$ mechanism.

\vspace{5mm}

\subsection{Pulsations in stars}\label{sec2c}

Assuming that Cowling's approximation (the perturbation to the gravitational potential is negligible) is valid, and that the star structure varies slower with radius than the oscillation mode, it is possible to derive simple equations describing the pulsations of stars \cite{Shibahashi89}. In order to study the radial part $\tilde{\zeta}_r$ of the infinitesimal radial displacement

\begin{displaymath}
\zeta_r(r,\theta,\phi,t)=\sqrt{4\pi}~\tilde{\zeta}_r(r)Y_{\ell}^m(\theta,\phi)e^{-i\omega t},
\end{displaymath}

\noindent where $Y_{\ell}^m$ are the spherical harmonics and $\omega$ the pulsation of the mode, two quantities are of peculiar importance \cite{Christensen-Dalsgaard15}: the Brunt-V\"ais\"al\"a frequency $N_{\mathrm{BV}}$, defined as

\begin{equation}\label{bv}
N_{\mathrm{BV}}^2=g\left(\frac{1}{\Gamma_1 P}\frac{dP}{dr}-\frac{1}{\rho}\frac{d\rho}{dr}\right)\;\;\;\; \mathrm{where}\;\;\;\; \Gamma_1=\left.\frac{\partial P}{\partial \rho}\right|_{\mathrm{ad}}
\end{equation}           

\noindent and the Lamb frequency $S_{\ell}$, such that:

\begin{equation}\label{lamb}
S_{\ell}^2=\frac{\ell(\ell+1)}{r^2}c_s^2\;\;\;\; \mathrm{with}\;\;\;\; c_s=\sqrt{\left.\frac{\partial P}{\partial \rho}\right|_{\mathrm{S}}}.
\end{equation}

In the above equations (\ref{bv}) and (\ref{lamb}), $g$ is the gravity, $P$ the pressure, $S$ the entropy, $\rho$ the density $\Gamma_1$ the adiabatic (ad) gradient and $c_s$ the sound speed. The case where $|\omega|>N_{\mathrm{BV}}$ and $|\omega|>S_{\ell}$ corresponds to ``p'' (pressure) modes, which are acoustic waves dominated by pressure forces. If $|\omega|\gg N_{BV}$, we have

\begin{equation}
\frac{d^2\tilde{\zeta}_r}{dr^2}=-K(r)\tilde{\zeta}_r\;\;\;\; \mathrm{with}\;\;\;\; K(r)=\frac{\omega^2-S_{\ell}^2}{c_s^2},
\end{equation}

\noindent which means that ``p'' modes are stuck in the region located between the star radius and the turning point $r_t$ defined as

\begin{equation}
K(r)=0\rightarrow\omega=S_{\ell}\rightarrow r=r_t=\frac{\sqrt{\ell(\ell+1)}c_s}{\omega}.
\end{equation}

It is worth mentioning that sdB (subdwarf B) stars are subject to gravity (g) modes, which are also related to iron opacity.

The prerequisite for asteroseismological studies in that observations of the pulsations must have a sufficient time span (at least one rotation period) to ensure detection of a significant number of pulsation modes.

\subsection{$\beta$ Cephei-type stars and the $\kappa$ mechanism}\label{sec2d}

In 1902, Edwin Frost discovers the variability of the radial velocity (the photosphere approaches and receeds alternatively from the observer) of $\beta$ Cephei ($\beta$ Canis Majoris or Alfirk), which has a magnitude from +3.16 to +3.27 and a period of about 4.57 hours. Spectroscopists measure the effect of the radial velocity on the absorption lines from different levels of the star's atmosphere. The opacity of the ``iron group'' (Cr, Fe and Ni) is particularly important for the envelopes of $\beta$ Cephei-type stars (8 to 18 M$_{\odot}$). Such stars (for instance $\nu$ Eridani, $\gamma$ Pegasi, $\beta$ Crucis and $\beta$ Centauri) are hot blue-white stars of spectral class B, their temperature is $T\approx$ 200-300,000 K and their density $\rho\approx$ 10$^{-7}$-10$^{-6}$ g/cm$^3$. Their pulsations are driven by opacity and the acoustic modes are excited through the ``$\kappa$ mechanism'', consisting of the following steps:

\vspace{5mm}

(i) The inward motion of a layer tends to compress the layer and increase its density.

\vspace{5mm}

(ii) The layer becomes more opaque, the flux from the deeper layers gets stuck in the high-opacity region.

\vspace{5mm}

(iii) The heat increase causes a build-up of pressure pushing the layer back out again.

\vspace{5mm}

(iv) The layer expands, cools and becomes more transparent to radiation.

\vspace{5mm}

(v) Energy and pressure beneath the layer diminish.

\vspace{5mm}

(vi) The layer falls inward and the cycle repeats.

\vspace{5mm}

It is worth mentioning that ``Slowly Pulsating B'' (SPB) stars are subject to gravity modes (``g'' modes), also connected to iron opacity. Their mass is from 2 to 6 M$_{\odot}$.

In 2003, Aerts et al. detected the differential rotation (the core rotates faster than the surface) with depth of HD 129929 \cite{Aerts03}. The authors were also able to estimate the amount of overshooting in the convective core. In 2004, Ausseloos et al. mentioned that no standard B-star model can explain observed frequencies in $\nu$ Eridani \cite{Ausseloos04}. An increase of iron abundance perhaps throughout the star, is needed. It seems that mixed pulsating modes are reproduced by none of the opacity tables. Eight years later, Salmon et al. found that in the low-metallicity $\beta$ Cephei-type stars of the Magellanic cloud, increasing iron opacity would not solve the problem while nickel could \cite{Salmon12}. In 2016, Moravveji computed new opacity tables with enhanced Fe and Ni contributions to Rosseland mean by 75 \% and reproduced the observed position of 10 stars on the Kiel diagram \cite{Moravveji16}. In 2017, Dasz\'ynska-Daszkiewicz et al. concluded that OPLIB opacities were preferred over OPAL and OP \cite{Daszynska-Daszkiewicz17}. They also reported an enhancement of the efficiency of convection in the $Z$ bump, still for $\nu$ Eridani. Very recently, Hui-Bon-Hoa and Vauclair showed that atomic diffusion leads to an overabundance of the iron-peak elements in the upper part of the envelope of B-type stars. The opacities may become as high as required, provided that fingering mixing is taken into account \cite{Hui-Bon-Hoa18}.

\section{Computation of atomic data and radiative opacity}\label{sec3}

\subsection{The different processes for plasmas in local thermodynamic equilibrium}\label{sec3a}

The opacity is the sum of photo-excitation (bb: bound-bound), photo-ionization (bf: bound-free), inverse Bremstrahlung (ff: free-free), corrected by stimulated-emission effect $(1-e^{-\frac{h\nu}{k_BT}})$, and scattering (s) contributions:

\begin{equation}
\kappa(h\nu)=\left[\kappa_{\mathrm{bb}}+\kappa_{\mathrm{bf}}+\kappa_{\mathrm{ff}}\right](h\nu)\times(1-e^{-\frac{h\nu}{k_BT}})+\kappa_s(h\nu).
\end{equation}

\noindent Photo-excitation can be described as

\begin{equation}
X_i^{q+}+h\nu \rightarrow X_j^{q+},
\end{equation}

\noindent where $X_i^{q+}$ is an ion with charge $q$ in an excitation state $i$. The signature of the absorbed photon $h\nu$ is a spectral line. The corresponding opacity contribution reads 

\begin{equation}
\kappa_{\mathrm{bb}}(h\nu)=\frac{1}{4\pi\epsilon_0}\frac{\mathcal{N}_A}{A}\frac{\pi e^2h}{m_ec}\sum_{i\rightarrow j}\mathcal{P}_if_{i\rightarrow j}\Psi_{i\rightarrow j}(h\nu),
\end{equation}

\noindent where $\mathcal{P}_i$ is the population of initial level $i$, $f_{i\rightarrow j}$ the oscillator strength and $\Psi_{i\rightarrow j}$ the profile of the spectral line corresponding to the transition $i\rightarrow j$, accounting for broadening mechanisms (Doppler, Stark, ...). $\epsilon_0$ is the dielectric constant, $\mathcal{N}_A$ the Avogadro number, $e$ and $m_e$ represent respectively the electron charge and mass, $c$ the speed of light and $A$ the atomic mass of the considered element. Photo-ionization is a threshold process that occurs when a bound electron $e^-$ is ejected after absorption of a photon with a high enough energy:

\begin{equation}
X_i^{q+}+h\nu \rightarrow X_j^{(q+1)+}+e^-.
\end{equation}

Bremsstrahlung refers to the radiation emitted by an electron slowing down in the electromagnetic field of an ion. The inverse process occurs when a free electron and an ion absorb a photon 

\begin{equation}
h\nu+\left[X_i^{q+}+e^-(\epsilon)\right]\rightarrow X_j^{q+}+e^-\left(\epsilon'\right),
\end{equation}

\noindent $\epsilon$ and $\epsilon'$ being the energies of the free electron before and after the photo-absorption. Calculations of the free-free cross-section involve quantities related to the matrix elements of elastic scattering for electron-impact excitation of ions. Scattering of a photon by a free electron can be accounted for using Klein-Nishina differential cross-section and Rayleigh scattering by a bound electron can be modeled by Kramers-Heisenberg cross-section.

\subsection{SCO-RCG code}\label{sec3b}

The detailed (fine-structure) opacity code SCO-RCG \cite{Pain15} enables one to compute precise opacities for the calculation of accurate Rosseland means. The (super-)configurations are generated on the basis of a statistical fluctuation theory and a self-consistent computation of atomic structure is performed for each configuration. In such a way, each configuration has its own set of wavefunctions. The latter are determined in a single-configuration approximation, which means that the so-called ``general configuration interaction'' is not taken into account completely (only interaction between relativistic sub-configurations of a non relativistic configuration \cite{Gilles12,Gilles15,Gilles16}. One peculiarity of the code is that it does not rely on the ``isolated atom'' picture, but on a realistic atom-in-plasma modeling (equation of state). Relativistic effects are taken into account in the Pauli approximation. The Detailed-Line-Accounting part of the spectrum is performed using an adapted version of the RCG routine from Cowan's suite of atomic-structure and spectra codes \cite{Cowan81}. The RCG source code was used for decades by spectroscopists, it has many available options and is well documented. In SCO-RCG, criteria are defined to select transition arrays that can be treated line-by-line. The data required for the calculation of the detailed transition arrays (Slater, spin-orbit and dipolar integrals) are obtained from SCO, providing in this way a consistent description of the plasma screening effects on the wavefunctions. Then, the level energies and the lines are calculated by RCG. In cases where the numbers of configurations and/or lines are too large, the code resorts to statistical methods \cite{Bauche-Arnoult79,Bauche-Arnoult85,Bar89}.

\subsection{Mean opacity values}\label{sec3c}

The Planck mean opacity is the opacity averaged over the Planck function:

\begin{equation}
\kappa_P=\int_0^{\infty}W_P(u)\kappa(u)du\;\;\;\; \mathrm{with}\;\;\;\; W_P(u)=\frac{15u^3e^{-u}}{\pi^4}.
\end{equation}

The Rosseland mean opacity is the harmonic mean opacity averaged over the derivative of the Planck function with respect to the temperature:

\begin{equation}
\frac{1}{\kappa_R}=\int_0^{\infty}\frac{W_R(u)}{\kappa(u)}du\;\;\;\; \mathrm{with}\;\;\;\; W_R(u)=\frac{15}{4\pi^4}\frac{u^4e^u}{\left(e^u-1\right)^2}.
\end{equation}

Table \ref{tab:table_compa} shows the Rosseland mean opacities computed by OP, ATOMIC \cite{Colgan17} and SCO-RCG codes for conditions of stellar envelopes. We can see that the values SCO-RCG values are rather close to the ATOMIC ones, and that the OP values are always smaller. The discrepancy between OP and SCO-RCG (or ATOMIC) are more important for the lowest temperatures.

\begin{table}[t]
	\centering
    \caption{Rosseland mean opacities computed by OP, ATOMIC \cite{Colgan17} and SCO-RCG codes for conditions of stellar envelopes \cite{Turck16}.}
	\label{tab:table_compa}
	\begin{tabular*}{\linewidth}{c @{\extracolsep{\fill}} c c c c}
	\noalign{\smallskip}\hline\hline\noalign{\smallskip}
$T$ & $\rho$ & OP & ATOMIC & SCO-RCG\\
(eV) & (g.cm$^{-3}$)&  &  & \\
	\noalign{\smallskip}\hline\noalign{\smallskip}
10.8 & 1.35 10$^{-6}$ & 25 & 64 & 63\\
15.3 & 3.4 10$^{-6}$ & 358 & 683 & 674\\
17.2 & 9.5 10$^{-7}$ & 354 & 487 & 500\\
21.6 & 8.8 10$^{-6}$ & 1270 & 1359 & 1313\\
25.5 & 2.4 10$^{-6}$ & 232 & 131 & 122\\
	\noalign{\smallskip}\hline
	\end{tabular*}
\end{table}

Silicates are an important component of cosmic matter and form in the diffuse interstellar medium in the winds of AGB (Asymptotic Giant Branch) stars (evolved cool luminous stars). They are found also around Ae/Be stars, T Tauri stars and brown dwarfs and are present in dust of protoplanetary disks. We can see in Table \ref{tab:table_si} that at $T$=20 eV, the Planck and Roseland mean opacities increase significantly with the density. 

\begin{table}[t]
	\centering
    \caption{Planck and Rosseland mean opacities computed by SCO-RCG for silicon at $T$=20 eV.}
	\label{tab:table_si}
	\begin{tabular*}{0.85\linewidth}{c @{\extracolsep{\fill}} c c}
	\noalign{\smallskip}\hline\hline\noalign{\smallskip}
$N_e$ & $\kappa_P$ & $\kappa_R$ \\
(cm$^{-3}$) & (cm$^2$/g) & (cm$^2$/g)\\
	\noalign{\smallskip}\hline\noalign{\smallskip}
10$^{18}$ & 7070 & 24.7 \\
10$^{19}$ & 8830 & 125.1\\
10$^{20}$ & 11920 & 593.1\\
	\noalign{\smallskip}\hline
	\end{tabular*}
\end{table}

\section{Interpretation of spectroscopy experiments}\label{sec4}

High-power lasers and Z-pinches can be used to produce X-ray fluxes which volumetrically heat materials to substantial temperatures. In such experiments, also known as pump-probe experiments, this X-ray flux, which is expected to be Planckian (\emph{i.e.} close to a blackbody radiation), creates a state of high-energy density matter that can be studied by the technique of absorption spectroscopy. 

\subsection{Laser experiment}\label{sec4a}

\begin{figure*}[ht]
	\centering
	\includegraphics[width=0.45\linewidth]{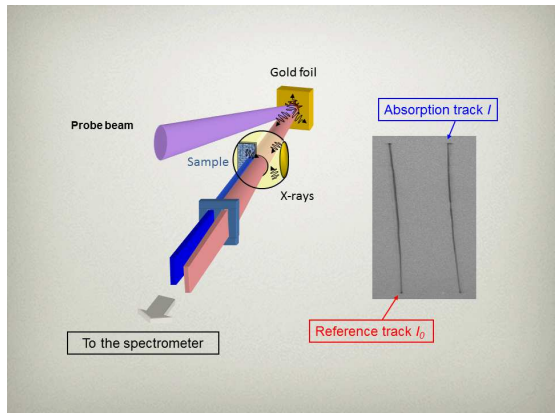}
	\caption{Example of experimental setup typical of absorption-spectroscopy measurements. $I$ is the attenuated signal, $I_0$ the reference one, and the ratio $I/I_0$ gives the transmission $T(h\nu)$ which is related to opacity by Beer-Lambert's law 
	$T(h\nu)=e^{-\rho L\kappa(h\nu)}$, where $\rho$ represents the matter density and $L$ the thickness of the sample.}
	\label{fig:setup}
\end{figure*}

In a laser experiment (see Fig. \ref{fig:setup}), laser beams are focused in a gold Hohlraum, their energy is converted into X rays heating a massive or a fiber target which becomes a plasma. A point-like X-ray source is created by tightly focusing a second laser beam on a foil, in order to obtain an intense continuum emission in the probed spectral range. 

\begin{figure*}[ht]
    \vspace{1cm}
	\centering
	\includegraphics[width=0.45\linewidth]{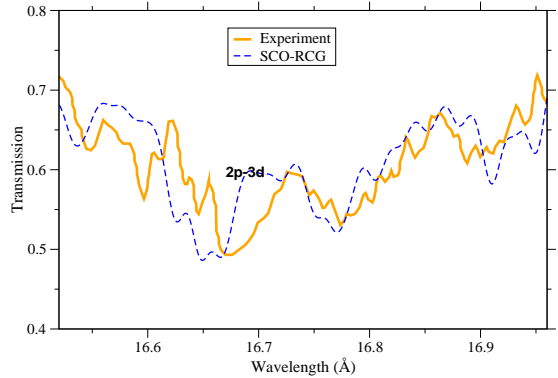}
	\caption{Comparison between an experimental transmission spectrum (in orange) of Fe measured on the ASTERIX IV laser facility \cite{Chenais00} and a SCO-RCG calculation (in dotted blue) at $T=20$~eV and $\rho=0.004$~g/cm$^{3}$. The areal mass is 8 $\mu$g/cm$^2$ and the resolving power $E/\Delta E=300$.}
	\label{fig:iron_chenais}
	\vspace{1cm}
\end{figure*}

Figure \ref{fig:iron_chenais} displays a comparison of the experimental and SCO-RCG calculated transmission spectra of Fe at $T$=20 eV and $\rho$=0.004 g/cm$^{3}$ measured on the ASTERIX IV laser facility in Germany \cite{Chenais00}. The spectrum corresponds to a part of a spectrum of the quasar IRAS 13349 + 2438 measured by XMM-Newton observatory of ions Fe VII-XII. Although the main absorption structures are reproduced, the agreement is not perfect, as concerns the transition energies and the transmission levels. The discrepancies might be attributed to temperature and density spatial and temporal non-uniformities, and to uncertainties in the knowledge of the areal mass. 

\begin{figure*}[ht]
	\centering
	\includegraphics[width=0.45\linewidth]{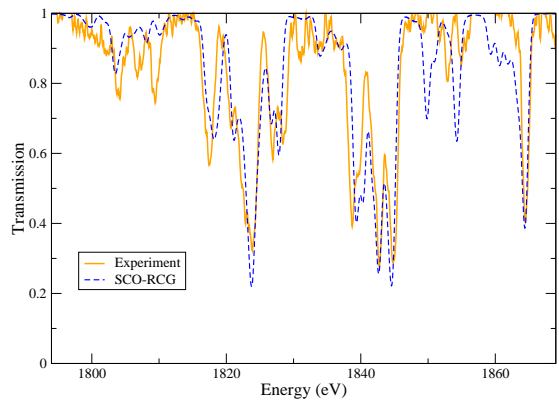}
	\caption{Comparison of the experimental (in orange) and calculated (in dotted blue) transmission spectra of Si at $T$=72 eV and $\rho$=0.006 g/cm$^{3}$ measured on the SG-II laser facility \cite{Xiong16}. The areal mass is 23 $\mu$g/cm$^2$ and the resolving power $E/\Delta E$=2000.}
	\label{fig:silicon}
\end{figure*}

In the past decades, the K-shell absorption lines of silicon ions in various astrophysical objects have been extensively observed with high-resolution spectrometers of the XMM-Newton, Chandra, and Suzaku space missions. Silicon absorption lines were observed in active galactic nuclei \cite{Xiong16}. The comparison between a recent experiment performed in China on ShenGuang-II (SG-II) (see Fig. \ref{fig:silicon}) for ions Si IX to Si XIII and SCO-RCG calculation shows a good agreement, except around $h\nu$=1855 eV. The differences are expected to be due to configuration interaction (between non-relativistic configurations, see Sec. \ref{sec3b}), not included in the computation (investigation is in progress).

\subsection{Z-pinch experiments}\label{sec4b}

\begin{figure*}[ht]
	\centering
	\includegraphics[width=0.60\linewidth]{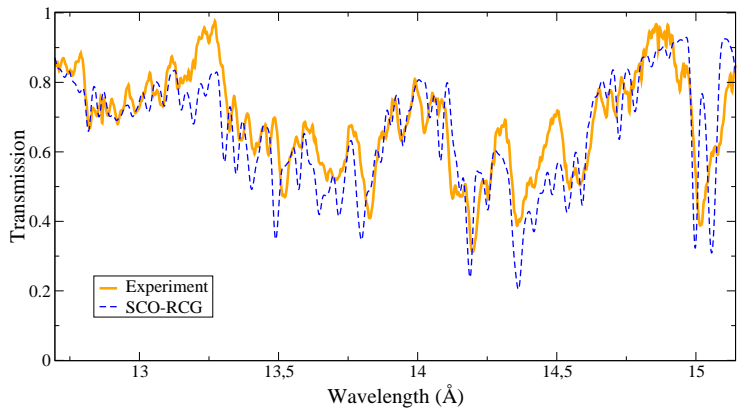}
	\caption{Comparison between the experimental spectrum (in orange) of \cite{Bailey07} and the SCO-RCG calculation (in dotted blue) at $T= 150$~eV and $\rho=0.058$~g/cm$^{-3}$.}
	\label{fig:fitBai}
	\vspace{1cm}
\end{figure*}

In 2007, Bailey \emph{et al.} reported on iron transmission measurements at $T$= 156 eV and $N_e$=6.9$\times$10$^{21}$ cm$^{-3}$ over the photon energy range $h\nu\approx$ 800-1800 eV \cite{Bailey07}. The samples consisted of an Fe/Mg mixture, fabricated by depositing 10 alternating Mg and Fe layers, fully tamped on both sides by a 10 $\mu m$ thick parylene-N (C$_8$H$_8$). The challenges of high-temperature opacity experiments were overcome here using the dynamic hohlraum X-ray source at the Sandia National Laboratory (SNL) Z-pinch facility. The process entails accelerating an annular tungsten Z-pinch plasma radially inward onto a cylindrical low density CH$_2$ foam, launching a radiating shock propagating toward the cylinder axis. Radiation trapped by the tungsten plasma forms a hohlraum and a sample attached on the top diagnostic aperture is heated during $\approx$ 9 ns when the shock is propagating inward and the radiation temperature rises above 200 eV. The radiation at the stagnation is used to probe the sample. The experimental spectrum was well reproduced by many fine-structure opacity codes (see Fig. \ref{fig:fitBai}).

\section{Boundary of the convective zone of the Sun}\label{sec5}

Iron contributes for 25 \% of the total opacity at the BCZ (boundary of the convective zone) of the Sun. Recent reevaluation of the abundances of C, N and O (see Table \ref{tab:table_soleil}) in the solar mixture \cite{Grevesse09} enhanced the disagreement between heliosismic measurements and predictions of the SSM (Standard Solar Model). In order to reconcile observations and modeling, a 5 to 20 \% increase would be necessary.

\begin{table}[t]
	\centering
    \caption{Solar abundances $A_X=\log10(N_X/N_H)+12$, where $N_X$ is the number of atoms of element $X$ (H, He, C, N, O et Fe).}
	\label{tab:table_soleil}
	\begin{tabular*}{0.85\linewidth}{c @{\extracolsep{\fill}} c c}
	\noalign{\smallskip}\hline\hline\noalign{\smallskip}
Element & Previous & New \\
        & abundances & abundances\\
	\noalign{\smallskip}\hline\noalign{\smallskip}
H & 12 & 12\\
He & 10.93 & 10.93\\
C & 8.52 & 8.43\\
N & 7.92 & 7.83\\
O & 8.83 & 8.69\\
Fe & 7.50 & 7.50\\
	\noalign{\smallskip}\hline
	\end{tabular*}
\end{table}

\begin{figure*}[ht]
	\centering
	\includegraphics[width=0.45\linewidth]{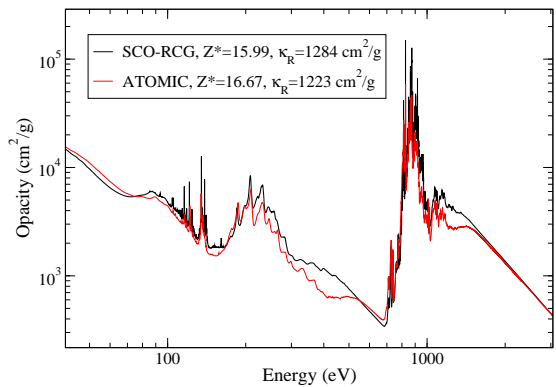}
	\caption{Comparison between SCO-RCG (in black) and ATOMIC (in red) in conditions close to the ones of the BCZ.}
	\label{fig:bcz}
\end{figure*}

Figure \ref{fig:bcz} shows that the opacity of iron at the BCZ are very close to eachother. The Rosseland means are respectively 1284 cm$^2$/g for SCO-RCG and 1223 cm$^2$/g for ATOMIC \cite{Colgan17}, which corresponds to a relative difference of about 5 \%. The OP Rosseland mean in these conditions is much lower and equal to 854 cm$^2$/g.

The iron opacity recently measured on the Z-pinch machine \cite{Bailey15} at temperature $T$=182 eV and electron density $N_e$=3.1$\times$10$^{22}$ cm$^{-3}$ (conditions close to the BCZ) is twice higher than all the theoretical spectra \cite{Pain15,Pain17,Pain18a}. The unexplained iron experiment on the SNL Z-pinch machine stimulates new developments in the codes. Recently, Nahar \emph{et al.} reported extensive R-matrix calculations \cite{Nahar11,Nahar16} of unprecedented complexity for iron ion Fe XVII, with a wavefunction expansion of 99 Fe XVIII LS core states from $n\leq 4$ complexes (equivalent to 218 fine-structure levels) and found a large enhancement in background photo-ionization cross-sections (up to orders of magnitude) in addition to strongly peaked photo-excitation-of-core resonances. However, as can be seen in Table \ref{tab:table_nahar}, the resulting Rosseland mean (NP2016) is now comparable to the SCO-RCG one, computed within the so-called ``distorted-wave'' formalism \cite{Blancard16,Delahaye18}.

\begin{table}[t]
	\centering
	\caption{Comparison of the Fe XVII Rosseland mean opacity (in cm$^2$/g) from different calculations \cite{Blancard16,Delahaye18}. The OP release is taken as reference.}
	\label{tab:table_nahar}
	\begin{tabular*}{0.85\linewidth}{l @{\extracolsep{\fill}} c l}
	\noalign{\smallskip}\hline\hline\noalign{\smallskip}
 Method & $\kappa_R\times$ 0.196 (cm$^2$/g) & $\kappa_R$/OP \\
	\noalign{\smallskip}\hline\noalign{\smallskip}
OP & 126.06 & 1.00\\
NP2016 & 170.18 & 1.35\\
ATOMIC & 166.40 & 1.32\\
OPAS & 195.39 & 1.55\\
SCO-RCG & 172.70 & 1.37\\
SCRAM & 160.10 & 1.27\\
TOPAZ & 152.53 & 1.21\\
	\noalign{\smallskip}\hline
	\end{tabular*}
\end{table}

\begin{figure*}[ht]
    \vspace{1cm}
	\centering
	\includegraphics[width=0.45\linewidth]{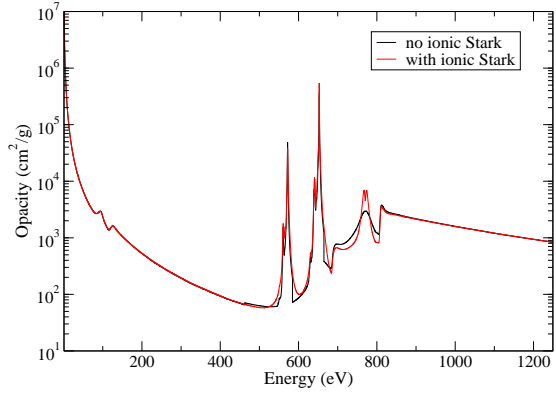}
	\caption{Opacity of oxygen at $T$=192.92 eV and $N_e$=10$^{23}$ cm$^{-3}$ (conditions close to the BCZ) with (in red) and without (in black) Stark effect.}
	\label{fig:oxygen}
\end{figure*}

Iron is not the only important element in the Sun. Lighter elements, such as oxygen, neon or fluorine also play a very important role. In that case, opacity is governed by absorption in the K shell, and in that case, the number of lines is small, but the Rosseland mean opacity is very sensitive to the line shapes, and especially the ionic Stark effect (see Fig. \ref{fig:oxygen} and Table \ref{tab:table_stark}).

\begin{table}[t]
	\centering
	\caption{Planck and Rosseland means in the case of an oxygen plasma at $T$=192.91 eV, $N_e=10^{23}$ cm$^{-3}$, conditions of the base of the convective zone of the Sun.}
	\label{tab:table_stark}
	\begin{tabular*}{0.85\linewidth}{l @{\extracolsep{\fill}} c l}
	\noalign{\smallskip}\hline\hline\noalign{\smallskip}
 & $\kappa_P$ (cm$^2$/g) & $\kappa_R$ (cm$^2$/g)\\
	\noalign{\smallskip}\hline\noalign{\smallskip}
No ionic Stark & 1685 & 273.5\\
With ionic Stark & 1719 & 285.3\\
	\noalign{\smallskip}\hline
	\end{tabular*}
\end{table}

\section{Possible role of two-photon processes in radiative accelerations}\label{sec6}

R. M. More and J.-C. Pain are also trying to figure out whether two-photon opacity \cite{More17,Pain18b} could bring some elements of explanation for the iron experiment. In addition, they are wondering whether two-photon absorption affects radiative diffusion of high-$Z$ ions in stars. Radiative diffusion is suppressed in convection regions and affects weakly the photospheric abundances because the abundance inhomogeneities in the radiative zone reflect very attenuated in the photosphere. However, it constitutes a cornerstone of stellar structure and evolution \cite{Michaud70}. 

\begin{figure*}[ht]
	\centering
	\includegraphics[width=0.85\linewidth]{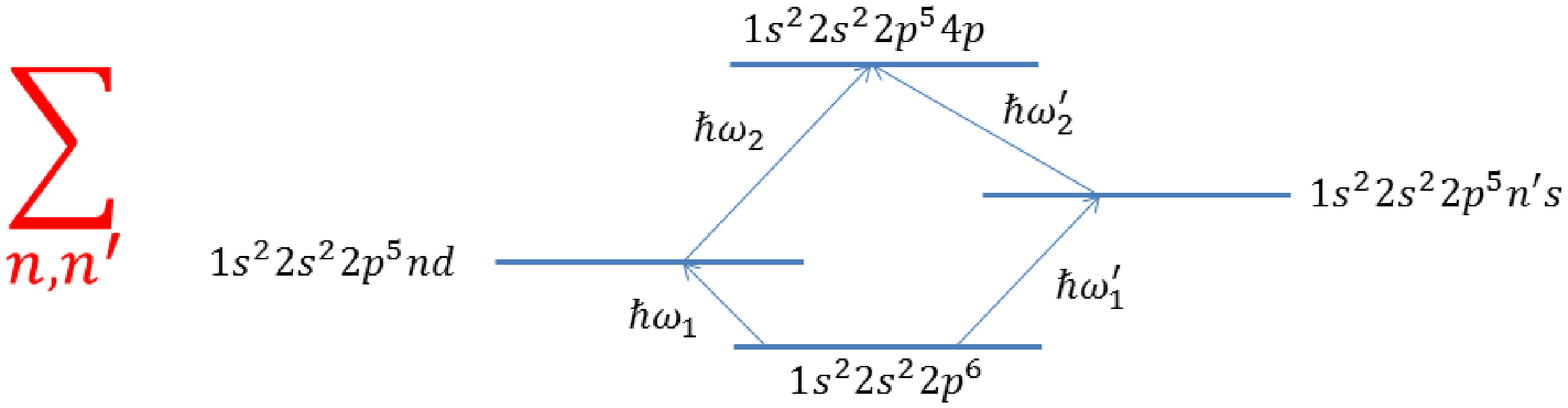}
	\caption{Example of two-photon processes. The initial and final configurations are connected by two families of channels, characterized respectively by intermediate states $1s^22s^22p^5nd$ and $1s^22s^22p^5n's$. The summations run over principal quantum number $n$ (for the first group) and $n'$ (for the second group).}
	\label{fig:twophoton}
\end{figure*}

The problem of acceleration (or diffusion) of ionic species in star interiors \cite{Vauclair77,Turcotte98}, induced by two-photon radiation was, as far as we know, never studied before. Astrophysicists usually take opacity tables (mostly OP and OPAL) as an input of their radiative-acceleration calculations. In order to compute the contribution of two-photon processes (see an example in Fig. \ref{fig:twophoton}), one has to sum over all possible intermediate states. The diffusion current is related to the opacity as an integral over the ``out-of-equilibrium'' part of the radiation field. For two-photon absorption, there is a similar integral, a little more complicated, which might yield extra diffusion due to the two-photon absorption. In particular, a phenomenon might be affected: the so-called ``saturation effect'' \cite{Alecian00}. When matter density increases,the number of ions per volume unit getting higher, the number of available photons likely to yield the acceleration decreases. For two-color absorption, saturation should be weaker, because of the many possible intermediate states, and the process is less stringent as concerns the photon energy. We are performing calculations to figure out whether there is a strong two-photon opacity effect. If the experimental opacity (for Fe) is higher than predicted, that should increase the radiative diffusion compared to the standard calculations, whatever the cause.

\clearpage

\section{Conclusion, work in progress and perspectives}\label{sec7}

As pointed out recently \cite{Lynas-Gray18}, there is a strong need for detailed accurate and complete energy levels, line positions, oscillator-strength values and cross-sections for atoms/ions/anions/molecules and transient ion dipoles formed in collisions. Some of the important topics which deserve particular attention and efforts are for instance:

\vspace{5mm}

(i) The accuracy / completeness compromise, which should be eased by the steady increase in computing power.

\vspace{5mm}

(ii) Plasma density effects: pressure ionization (quasi-bound states), realistic microfield distributions for reliable line broadening.

\vspace{5mm}

(iii) Stark and pressure (van der Waals) broadenings, broadened Fano profiles for autoionization, interference with spectator electrons. Line absorption is redistributed in frequency due to Stark effect, which is important for all relevant L- and M-shell lines.

\vspace{5mm}

(iv) Quantum-mechanical calculation of continuum photo-absorption: photo-ionization and inverse Bremsstrahlung. The continuity of oscillator strength must be ensured.

\vspace{5mm}

(v) Proper accouting for plasma oscillations on the dielectric constant \cite{Sarfraz18}, including electron degeneracy and screening.

\vspace{5mm}

Opacity computation requires to take into account a huge number of levels and spectral lines. Some effects, such as configuration interaction, are still difficult to take into account properly. Laser or Z-pinch experiments are performed in order to test the models. The unexplained iron experiment on the SNL Z-pinch machine yields new developments in order to figure out if the models are lacking for physicical processes (photo-ionization, highly-excited states, two-photon processes \cite{More17,Pain18b}, etc. In the same conditions, we will see whether the nickel spectrum is in good agreement with SCO-RCG or if the same disagreement exists as for iron. An experiment is ongoing on NIF laser facility in order to measure iron opacity in the same conditions as the Z-pinch SNL experiment \cite{Heeter18}. The SCO-RCG code is now ready (accuracy, completeness, robustness) to produce opacity tables. We are also currently developing a new lineshape code \cite{Gilleron18}, named ZEST (ZEeman-STark), which should hopefully be useful for many astrophysical applications, for instance to interpret the measurements of H$_{\beta}$, H$_{\gamma}$, and H$_{\delta}$ lines at white dwarf photospheric / atmospheric conditions \cite{Falcon13}. It is very important to continue helioseismological work, in particular probing BCZ and He II ionization zone. Further study of pulsating stars, chemically peculiar stars and asteroseismology in the context of hybrid pulsators should be helpful in elucidating opacity shortcomings.

\section*{Acknowledgments}

J.-C. Pain would like to thank the organizers for the invitation to give a review talk on opacities and asteroseismology at the Physics of Oscillating Stars (PHOST) conference (Banyuls-sur-Mer, France, 3-7 September 2018). The conference was a great opportunity to forge links with the best specialists in asteroseismology and to better guide the work on opacity calculation for astrophysical applications.

\end{document}